\documentclass[aps,prl,twocolumn,superscriptaddress]{revtex4-1}
\usepackage{graphics}
\usepackage{graphicx}
\usepackage{float}
\usepackage{amsmath}
\usepackage{color}
\usepackage{verbatim}
\usepackage[normalem]{ulem}

\begin{document}
\title{Capillary levelling of free-standing liquid nanofilms}
\author{Mark Ilton}
\affiliation{Department of Physics \& Astronomy, McMaster University, Hamilton, Ontario, Canada, L8S 4M1}
\author{Miles M. P. Couchman}
\affiliation{Department of Physics \& Astronomy, McMaster University, Hamilton, Ontario, Canada, L8S 4M1}
\author{Cedric Gerbelot}
\affiliation{Laboratoire de Physico-Chimie Th\'{e}orique, UMR CNRS Gulliver 7083, ESPCI Paris, PSL Research University, 75005 Paris, France}
\author{Michael Benzaquen}
\affiliation{Laboratoire de Physico-Chimie Th\'{e}orique, UMR CNRS Gulliver 7083, ESPCI Paris, PSL Research University, 75005 Paris, France}
\author{Paul D. Fowler}
\affiliation{Department of Physics \& Astronomy, McMaster University, Hamilton, Ontario, Canada, L8S 4M1}
\author{Howard A. Stone}
\affiliation{Department of Mechanical and Aerospace Engineering, Princeton University, Princeton,
	New Jersey 08544, USA}
\author{Elie Rapha\"{e}l}
\affiliation{Laboratoire de Physico-Chimie Th\'{e}orique, UMR CNRS Gulliver 7083, ESPCI Paris, PSL Research University, 75005 Paris, France}
\author{Kari Dalnoki-Veress}
\affiliation{Department of Physics \& Astronomy, McMaster University, Hamilton, Ontario, Canada, L8S 4M1}
\affiliation{Laboratoire de Physico-Chimie Th\'{e}orique, UMR CNRS Gulliver 7083, ESPCI Paris, PSL Research University, 75005 Paris, France}
\author{Thomas Salez}
\email{thomas.salez@espci.fr}
\affiliation{Laboratoire de Physico-Chimie Th\'{e}orique, UMR CNRS Gulliver 7083, ESPCI Paris, PSL Research University, 75005 Paris, France}
\affiliation{Department of Mechanical and Aerospace Engineering, Princeton University, Princeton,
New Jersey 08544, USA}
\affiliation{Global Station for Soft Matter, Global Institution for Collaborative Research and Education, Hokkaido University, Sapporo, Hokkaido 060-0808, Japan}
\date{\today}
\begin{abstract}
We report on the capillary-driven levelling of a topographical perturbation at the surface of a free-standing liquid nanofilm. The width of a stepped surface profile is found to evolve as the square root of time. The hydrodynamic model is in excellent agreement with the experimental data. In addition to exhibiting an analogy with diffusive processes, this novel system serves as a precise nanoprobe for the rheology of liquids at interfaces in a configuration that avoids substrate effects.
\end{abstract}

\maketitle
Continuum fluid dynamics provides a remarkably accurate description of flows from nanometric to astronomical length scales, and can accommodate a variety of forces and effects such as inertia, gravity, surface tension, viscosity, etc. Capillary-driven flows mediated by viscosity represent a particularly interesting situation as these two mechanisms dominate at small length scales and thus in several biological systems and technologically relevant applications~\cite{Tabeling2005,Jacobs2008}. They also underlie conceptually simple experiments that probe physics at interfaces and in confinement, such as the motion of molecules in a liquid near a solid boundary~\cite{Baumchen2010b}, or the universal coalescence of droplets~\cite{Hernandez2012}. Moreover, the thin-film geometry allows for an important simplification since the in-plane flow dominates~\cite{Blossey2012}. In this so-called lubrication approximation, the dynamics is solely controlled by the ratio of the film-air surface tension to viscosity, known as the capillary velocity $v_{\textrm{c}}=\gamma/\eta$, and the film thickness which sets the characteristic length scale. 

Until now, research has been largely focused on thin films \textit{supported} on solid substrates~\cite{Oron1997,Craster2009}. One way to probe the dynamics of such a film is to apply an external stress and to measure the departure from the initial state. Measurements can be achieved through a variety of experimental techniques, such as the dynamic surface forces apparatus~\cite{Israelachvili1988}, including substrate elasticity~\cite{Villey2013}, nanoparticle embedding~\cite{Teichroeb2003}, electrohydrodynamic instability~\cite{Schaffer2000,Morariu2003,Voicu2006}, unfavorable wetting conditions~\cite{Srolovitz1986,Wyart1990,Seemann2001a,Reiter2005a,Chen2012,Baumchen2010b,Baumchen2009}, and Marangoni flow~\cite{Katzenstein2012,BinKim2014}. Alternatively, the dynamics of a thin coating of liquid can be probed by starting with an out-of-equilibrium surface topography and observing the film as it relaxes towards equilibrium~\cite{Tsui2008a,Fakhraai2008,Leveder2008,Yang2010,McGraw2012,Rognin2012,Chai2014}. All the above approaches can be used to study physical properties such as the glass transition~\cite{Teichroeb2003,Yang2010,Fakhraai2008,Chai2014}, viscoelasticity~\cite{Rognin2012}, and interfacial molecular friction~\cite{Baumchen2009,Baumchen2010b}. 

In contrast to these studies, one could examine capillary-driven flows mediated by viscosity in a configuration with no substrate effect: a \textit{free-standing} film. Flow in such a geometry has been previously studied for liquids with some internal structure that stabilizes the film against rupture, such as soap films~\cite{Rutgers1996,Horvath2000,Aradian2001,Georgiev2002,Diamant2009,Seiwert2013,Huang2013}, or liquid-crystal films~\cite{Harth2011,Eremin2006,Stannarius2006}, but this structure has necessarily a significant impact on the dynamics. Designing homogeneous isotropic free-standing liquid nanofilms presents experimental challenges in their creation, stability, and observation, as the techniques used for supported films typically do not work. Nevertheless, this geometry remains highly compelling as a way to study fundamental phenomena without substrate-induced artifacts. For instance, the glass-transition temperature reductions observed for supported films~\cite{Keddie1994} are more pronounced for free-standing films~\cite{Forrest1996}. Also, due to the absence of any liquid-substrate interaction, confinement effects of polymeric liquids~\cite{Silberberg1982,Brochard2000,Bodiguel2006,Shin2007} may be addressed in an ideal way using free-standing films~\cite{Dalnoki2000,Si2005}. Furthermore, the interfaces of a free-standing film are the perfect realization of infinite-slip-length boundaries, and their behaviour can shed light on theoretical predictions~\cite{Eggers1997,Kargupta2004,Munch2005,Howell2005}. Finally, this technique has the potential to resolve questions such as the influence of thermal noise at interfaces~\cite{Davidovitch2005,Grun2006,Hennequin2006}.

In this Letter, we report on a novel capillary-levelling experiment involving free-standing stepped nanofilms. First, using atomic force microscopy (AFM), we empirically determine the scaling law for the width evolution of the prepared samples. Then, we derive a hydrodynamic model showing excellent agreement with experimental data. The only adjustable parameter is the capillary velocity, which allows the system to be used as a rheological nanoprobe. Finally, by performing simultaneous experiments on both supported and free-standing films, we self-consistently confirm the robustness of the technique.
\begin{figure}[t!]
\includegraphics[width=80mm]{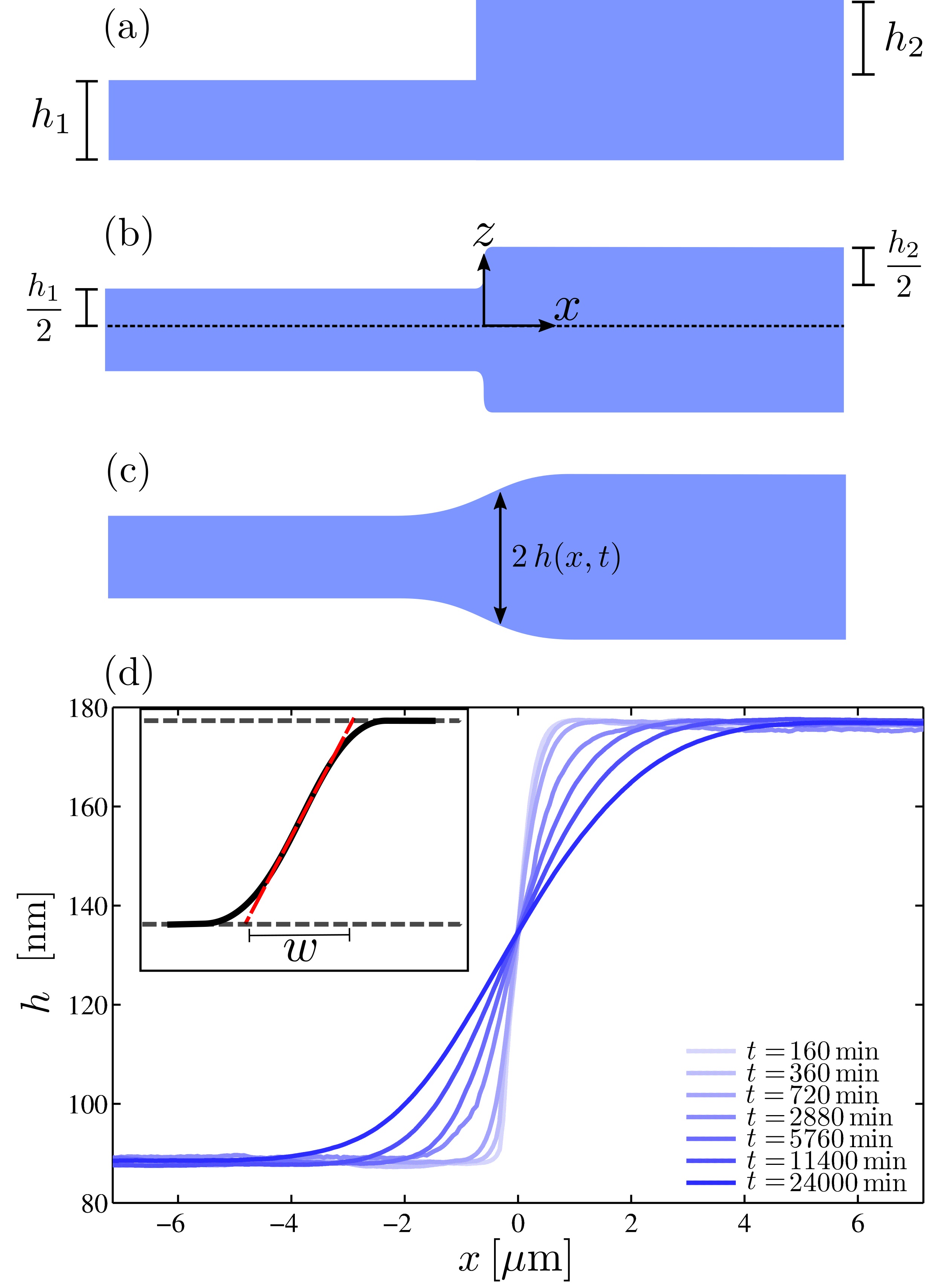}
\caption{(a) Schematic of a free-standing stepped film. (b) In the melt state, the profile quickly symmetrizes with respect to the central plane (dotted line) at $z=0$. (c) The two mirrored steps broaden horizontally over time. The half-height profile $h(x,t)$ is recorded with AFM. (d) AFM profiles of a free-standing stepped film with $h_1=h_2=176\,\textrm{nm}$, for different times $t$ spent above $T_{\textrm{g}}$. As sketched in the inset (arbitrary units), the profile width $w$ is defined from the tangent (tilted dashed) line in the middle of the step and the film thickness.}
\label{Fig1}
\end{figure}

Free-standing stepped films are prepared using a protocol inspired in part from~\cite{McGraw2011}. Here, thin polystyrene (PS) films, with thickness in the $100-500\,\textrm{nm}$ range, are spin-coated from dilute PS (Polymer Source, Canada, with weight-averaged molecular weight 55 kg/mol) solution in toluene onto freshly cleaved mica (Ted Pella, USA), and pre-annealed for at least ten minutes on a hot stage (Linkham, UK) at $140\,^{\circ}\textrm{C}$, well above the glass-transition temperature $T_{\textrm{g}} \approx 100\,^{\circ}\textrm{C}$ of the material. After annealing, the films are floated onto a deionized water bath (18.2 M$\Omega$ cm, Pall, USA) and picked up onto custom-machined stainless steel grids. 
Each grid has 85 hexagonal holes ($\sim1\,\textrm{mm}$ across) that allow the simultaneous preparation of $\sim85$ isolated free-standing samples. All results described here are obtained from the few films that remain stable against spontaneous rupture over the course of the entire experiment. After floating a film with thickness $h_1$ onto the grid, it is heated above $T_{\textrm{g}}$ to remove wrinkles. A second film with thickness $h_2$ and a sharp edge~\cite{Baumchen2013} is then floated atop the first film, as schematically shown in Fig.~1(a). The free-standing stepped film is then annealed at $110\,^{\circ}\textrm{C}$ on the hot stage for a time $t$. To vertically equilibrate the pressure, the film rapidly symmetrizes into two mirrored steps as depicted in Fig.~1(b). Because the vertical length scale is orders of magnitude smaller than the typical horizontal one, the vertical symmetrization occurs rapidly on experimental time scales and is not resolved. 

The presence of an excess interfacial area with respect to the flat equilibrium state drives flow, thus causing the profile to broaden laterally over time, as shown schematically in Fig.~1(c). After a chosen evolution time, each film is then cooled rapidly ($>90\,^{\circ}\textrm{C/min}$) from $110\,^{\circ}\textrm{C}$ to room temperature, deep into the glassy state, and the profile is imaged using AFM (Veeco Caliber, USA), with all sources of vibrations minimized. Then, the sample is placed back on the hot stage and rapidly heated above $T_{\textrm{g}}$ to continue the levelling process. Because appreciable flow does not occur at room temperature, the film can be intermittently imaged this way.

Previous experiments on stepped films supported on a substrate~\cite{McGraw2012} have shown that the width $w$ of the profile increases with a $1/4$ power law in time:
\begin{align}\label{eq:wt14}
w \propto \left(v_{\textrm{c}} h_2^3 t\right)^{1/4}\quad \textrm{[supported film]}\ ,
\end{align}
where $h_2$ is the thickness of the top film as in Fig~1(a), and where the missing numerical prefactor depends on the aspect ratio $h_2/h_1$. Equation~(\ref{eq:wt14}) is obtained from the capillary-driven thin-film equation~\cite{Oron1997,Craster2009,Blossey2012}, which results from the incompressible Stokes equation in the lubrication approximation, together with a no-slip boundary condition at the solid-liquid interface and no shear stress at the liquid-air interface. In that case, there is a nearly unidirectional flow with a parabolic velocity profile within the film.

For the free-standing stepped-film experiments studied here, a typical temporal evolution is shown in Fig.~1(d). To characterize the broadening, the width $w$ of the profile~\cite{McGraw2011} is defined (Fig.~1(d), inset) and recorded as a function of time. As seen in Fig.~2, the evolution is consistent with a $t^{1/2}$ power law. Also shown are the results for two other thicknesses (keeping $h_1=h_2$). Each dataset is then fit to $w = (M t)^{1/2}$, where the mobility $M$ is a free parameter that is found to scale linearly with $h_2$ (Fig.~2, inset). Using dimensional analysis, this empirically gives:
\begin{align}\label{eq:wt12}
w \propto \left(v_{\textrm{c}} h_2 t\right)^{1/2}\quad \textrm{[free-standing film]}\ ,
\end{align}
where the missing prefactor is expected to depend on $h_2/h_1$. This scaling is similar to Eq.~(\ref{eq:wt14}), but with a different power-law exponent. There is a stronger dependence on time in free-standing films compared to supported films, which is consistent with the former flowing more easily due to the unconstrained boundaries.
\begin{figure}[t!]
\includegraphics[width=80mm]{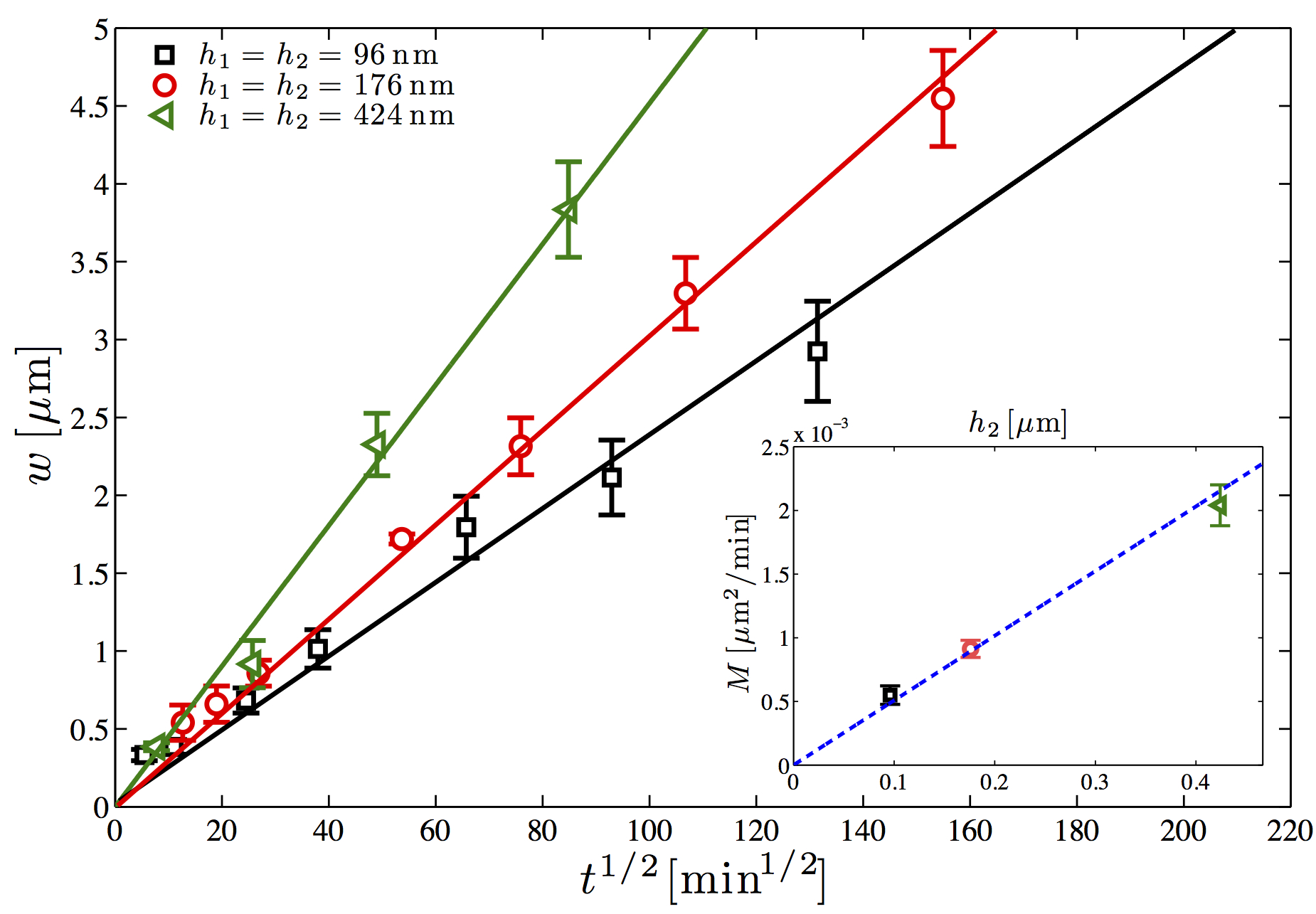}
\caption{Profile width $w$ (Fig.~1(d), inset) as a function of the square root of time $t^{1/2}$, for free-standing stepped films with $h_1=h_2$. The open symbols are the measured widths for three different film thicknesses, as indicated. The solid lines are fits to $w = (Mt)^{1/2}$, where the mobility $M$ is the fitting parameter. (inset) Mobility as a function of thickness $h_2$, when $h_1=h_2$. The dashed line is a linear fit.}
\label{Fig2}
\end{figure}

Motivated by the empirical scaling of Eq.~(\ref{eq:wt12}), we now turn to the theoretical description. Let us consider a symmetric free-standing film of total thickness $2 \, h(x,t)$ (Fig.~1(c)) that viscously flows under the action of liquid-air surface tension. According to previous research on free-standing geometries~\cite{Eggers1997,Howell2005} that invoke long-wave theory, the Young-Laplace equation in the small-slope approximation, as well as a no-shear boundary condition at each liquid-air interface, the mass and momentum conservations, respectively, lead to two coupled nonlinear equations:
\begin{eqnarray}
\label{pde1}
h_t+(hu)_x&=&0\\
\label{pde2}
\frac4{v_{\textrm{c}}}(hu_x)_x+hh_{xxx}&=&0\ ,
\end{eqnarray}
where $u(x,t)$ is the horizontal component of the fluid velocity, and where the subscripts $\cdot_x$ and $\cdot_t$ indicate the corresponding partial derivatives. Note the equivalence of these equations with the ones describing the infinite-slip-length flow in supported films of thickness $h(x,t)$~\cite{Kargupta2004,Munch2005}, in the absence of disjoining pressure.

\begin{figure}[t!]
\includegraphics[width=80mm]{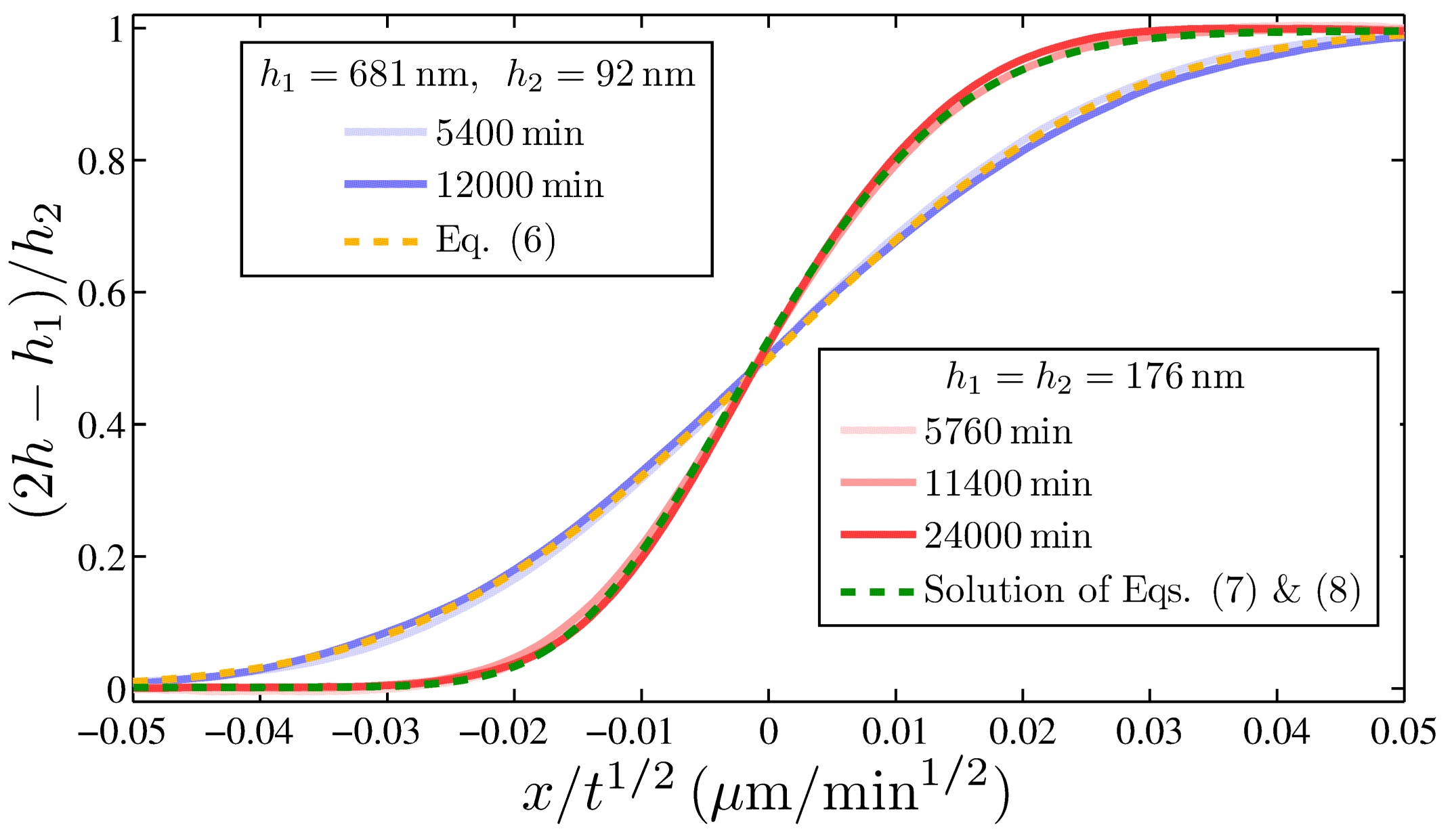}
\caption{Normalized long-time experimental height profiles as a function of the self-similar variable $x/t^{1/2}$, at several times and for two different geometries as indicated. The light/yellow dashed line represents a fit using Eq.~(\ref{eq:LFSTFE}). The dark/green dashed line indicates a fit using the numerical solution of Eqs.~(\ref{eq:FSTFE1}) and~(\ref{eq:FSTFE2}) for the $h_1=h_2$ case.}
\label{Fig3}
\end{figure}

Let us linearize Eqs.~(\ref{pde1}) and~(\ref{pde2}) around the flat equilibrium state characterized by $h=h_{\textrm{eq}}=h_1/2+h_2/4$ and $u=u_{\textrm{eq}}=0$, where we assume that $h_2/4 << h_{\textrm{eq}}$. 
Doing so, one gets a diffusive-like free-standing thin-film equation for the film thickness:
\begin{equation}
\label{diflin}
h_t=\frac{v_{\textrm{c}}}{4}\left(\frac{h_1}{2}+\frac{h_2}{4}\right)h_{xx}\ .
\end{equation}
Equation~(\ref{diflin}) can be solved analytically with the boundary conditions of Fig.~1(b), to arrive at:
\begin{equation}
h(x,t)=\frac{h_1}{2}+\frac{h_2}{4}\left[1+\textrm{erf}\left(\frac{2x}{\sqrt{\left(2h_1+h_2\right)v_{\textrm{c}}t}}\right)\right]\ .
\label{eq:LFSTFE}
\end{equation}
As seen in Fig.~\ref{Fig3}, Eq.~(\ref{eq:LFSTFE}) fits the experimental data remarkably well for a sample with $h_1=681\,\textrm{nm}$ and $h_2 =92\,\textrm{nm}$, and the $x/t^{1/2}$ self-similarity is confirmed by the collapse of the profiles at two different times.
The single fit parameter $v_{\textrm{c}}$ will be studied in detail below.

We now turn to the general nonlinear case. Let us first nondimensionalize Eqs.~(\ref{pde1}) and~(\ref{pde2}) through: $h=Hh_1/2$, $x=Xh_1/2$, $u=Uv_{\textrm{c}}/4$, and $t=2Th_1/v_{\textrm{c}}$. Inspired by the linear case above, and by the empirical scaling of Eq.~(\ref{eq:wt12}), we make the following self-similarity ansatz~\cite{Barenblatt1996}: $H(X,T)=F(S)$, and $U(X,T)=Q(S)/T^{1/2}$, where $S=X/T^{1/2}$. Assuming further that $Q'$, $F'$, and $F''$ vanish as $S\rightarrow-\infty$, while by construction $\lim_{S\rightarrow-\infty}F=1$, Eqs.~(\ref{pde1}) and~(\ref{pde2}) become:
\begin{eqnarray}
\label{eq:FSTFE1}
Q'&=&\frac{SF'}{2F}-\frac{QF'}{F}\\
\label{eq:FSTFE2}
F''&=&\frac{F'^2}{2F}-Q'\ .
\end{eqnarray}
We solve these equations using a one-parameter shooting method. In fact, as $S\rightarrow-\infty$, Eqs.~(\ref{eq:FSTFE1}) and~(\ref{eq:FSTFE2}) can be linearized around the boundary conditions $F=1$ and $Q=0$. The approximate far-field solution is $F\simeq1+c[1+\textrm{erf}(S/2)]$, together with $Q=-F'$, where $c$ is an unknown constant. This analysis allows us to evaluate $F$ and $Q$ at the left boundary of the integration domain ($-S\gg1$). For an arbitrary $c$, one can thus integrate Eqs.~(\ref{eq:FSTFE1}) and~(\ref{eq:FSTFE2}) numerically using a Runge-Kutta scheme. We then shoot over $c$ until the numerical solution reaches the proper limit, $\lim_{S\rightarrow+\infty}F=1+h_2/h_1$, at the right boundary of the integration domain ($S\gg1$). Figure~3 shows the long-time experimental profiles of the $h_1=h_2=176\,\textrm{nm}$ geometry of Fig.~\ref{Fig1}(d). The normalized height is plotted against the self-similar variable $x/t^{1/2}$, for three different annealing times. The profiles collapse onto a single master curve -- thus confirming the self-similarity ansatz -- and are in full agreement with the numerical solution. The vertical discrepancy is less than $1\,\textrm{nm}$, even after the hundreds of hours of flow in the experiments, and there is a comparable level of agreement (not shown) for the two other $h_1=h_2$ geometries introduced in Fig.~\ref{Fig2}. 

A final self-consistency check is performed in order to ensure the accuracy on the single fit parameter $v_{\textrm{c}}$ extracted from the comparison with theory. As a material property of the film in contact with air, it should not depend on the presence of an additional substrate-film interface. Stepped PS films ($55\,\textrm{kg/mol}$, $h_1=h_2=390\,\textrm{nm}$) are thus prepared in free-standing and substrate-supported (on Silicon wafers or $\sim4 \mu\textrm{m}$-thick polysulfone films) configurations. Both are annealed side-by-side, at the same time, in a vacuum oven at $120\,^{\circ}\textrm{C}$. The profiles are measured after $735$ and $1485$ minutes to confirm self-similarity. Figure~4(a) shows the results for the supported case. The normalized height profile is plotted as a function of the self-similar variable $x/t^{1/4}$ of Eq.~(\ref{eq:wt14}). The numerical profile~\cite{Salez2012a} is fit to the experimental data with one single free parameter: $v_{\textrm{c}} = 0.068 \pm 0.010\,\mathrm{\mu m /min}$. Figure~4(b) shows the results for the free-standing case. The normalized height profile is plotted as a function of the self-similar variable $x/t^{1/2}$ of Eq.~(\ref{eq:wt12}). The numerical solution of Eqs.~(\ref{eq:FSTFE1}) and~(\ref{eq:FSTFE2}) described above is fit to the experimental data with one single free parameter: $v_{\textrm{c}} = 0.087 \pm 0.012\,\mathrm{\mu m /min}$. Note the difference between the profile shapes of both cases, reflecting the different orders in the governing equations. Together with the excellent fit qualities, the fact that both cases give the same capillary velocity within experimental error proves: i) the robustness of the free-standing stepped-film technique as a rheological nanoprobe; and ii) the validity of the hydrodynamic model. 
\begin{figure}[t!]
\includegraphics[width=75mm]{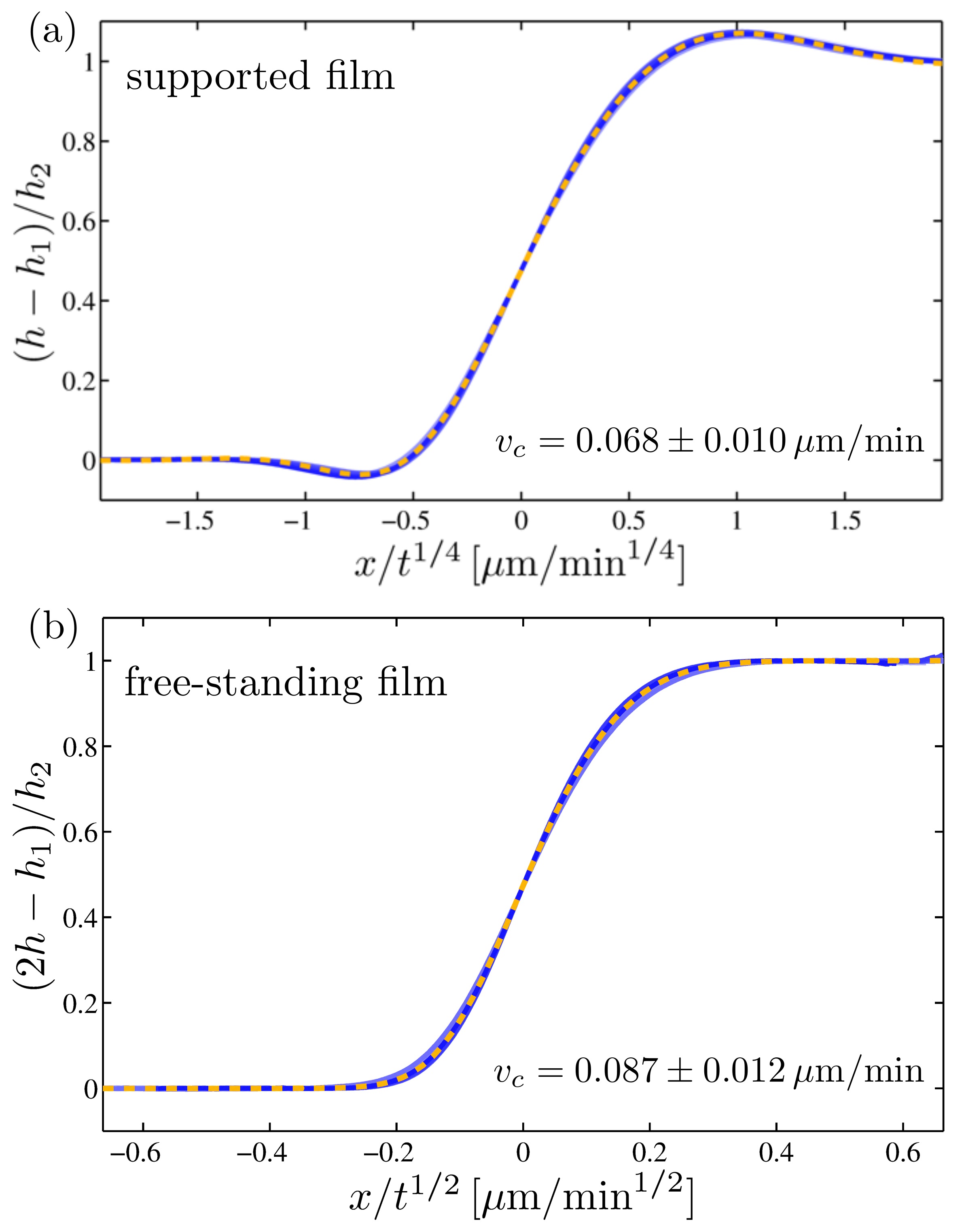}
\caption{(a) Normalized long-time experimental height profiles (solid lines) of supported PS stepped films ($55\,\textrm{kg/mol}$, $h_1=h_2=390\,\textrm{nm}$) as a function of the self-similar variable $x/t^{1/4}$ of Eq.~(\ref{eq:wt14}), for ten samples and two different times. The numerical solution~\cite{Salez2012a} is fit (dashed line) to the experimental profiles with $v_{\textrm{c}}$ as the free parameter. (b) Normalized long-time experimental height profiles (solid lines) of free-standing PS stepped films ($55\,\textrm{kg/mol}$, $h_1=h_2=390\,\textrm{nm}$) as a function of the self-similar variable $x/t^{1/2}$ of Eq.~(\ref{eq:wt12}), for twelve samples and two different times. The numerical solution of Eqs.~(\ref{eq:FSTFE1}) and~(\ref{eq:FSTFE2}), for the $h_1=h_2$ case, is fit (dashed line) to the experimental profiles with $v_{\textrm{c}}$ as the free parameter.}
\label{Fig4}
\end{figure}

In conclusion, we report on capillary-driven viscous flow in polystyrene free-standing stepped nanofilms. Above the glass-transition temperature, the surface profiles broaden with a characteristic $t^{1/2}$ power law, as measured by AFM. This response differs significantly from the $t^{1/4}$ power law of substrate-supported films with a no-slip liquid-solid boundary condition. The hydrodynamic model is found to be in excellent agreement with experimental data. Finally, the capillary velocity $\gamma/\eta$ is robustly and accurately extracted from the comparison with theory. We expect this novel system to be of fundamental interest for three reasons: first, it presents a striking analogy with diffusive processes; secondly, it embodies a perfect realization of the infinite-slip-length limit of low-Reynolds-number thin-film hydrodynamics and, as such, it is a test of the associated theoretical predictions; thirdly, by avoiding substrate-induced artifacts, it may serve as a precise nanoprobe for addressing fundamental questions about complex fluids and glass formers in confinement and at interfaces. 

\bigskip

The authors thank Martin Brinkmann and Ren\'e Ledesma-Alonso for interesting discussions, and gratefully acknowledge the financial support from NSERC of Canada and the Global Station for Soft Matter, a project of Global Institution for Collaborative Research and Education at Hokkaido University. 


\begin{thebibliography}{55}%
\makeatletter
\providecommand \@ifxundefined [1]{%
\@ifx{#1\undefined}
}%
\providecommand \@ifnum [1]{%
\ifnum #1\expandafter \@firstoftwo
\else \expandafter \@secondoftwo
\fi
}%
\providecommand \@ifx [1]{%
\ifx #1\expandafter \@firstoftwo
\else \expandafter \@secondoftwo
\fi
}%
\providecommand \natexlab [1]{#1}%
\providecommand \enquote  [1]{``#1''}%
\providecommand \bibnamefont  [1]{#1}%
\providecommand \bibfnamefont [1]{#1}%
\providecommand \citenamefont [1]{#1}%
\providecommand \href@noop [0]{\@secondoftwo}%
\providecommand \href [0]{\begingroup \@sanitize@url \@href}%
\providecommand \@href[1]{\@@startlink{#1}\@@href}%
\providecommand \@@href[1]{\endgroup#1\@@endlink}%
\providecommand \@sanitize@url [0]{\catcode `\\12\catcode `\$12\catcode
`\&12\catcode `\#12\catcode `\^12\catcode `\_12\catcode `\%12\relax}%
\providecommand \@@startlink[1]{}%
\providecommand \@@endlink[0]{}%
\providecommand \url  [0]{\begingroup\@sanitize@url \@url }%
\providecommand \@url [1]{\endgroup\@href {#1}{\urlprefix }}%
\providecommand \urlprefix  [0]{URL }%
\providecommand \Eprint [0]{\href }%
\providecommand \doibase [0]{http://dx.doi.org/}%
\providecommand \selectlanguage [0]{\@gobble}%
\providecommand \bibinfo  [0]{\@secondoftwo}%
\providecommand \bibfield  [0]{\@secondoftwo}%
\providecommand \translation [1]{[#1]}%
\providecommand \BibitemOpen [0]{}%
\providecommand \bibitemStop [0]{}%
\providecommand \bibitemNoStop [0]{.\EOS\space}%
\providecommand \EOS [0]{\spacefactor3000\relax}%
\providecommand \BibitemShut  [1]{\csname bibitem#1\endcsname}%
\let\auto@bib@innerbib\@empty
\bibitem [{\citenamefont {Tabeling}(2005)}]{Tabeling2005}%
\BibitemOpen
\bibfield  {author} {\bibinfo {author} {\bibfnamefont {P.}~\bibnamefont
{Tabeling}},\ }\href@noop {} {\emph {\bibinfo {title} {{Introduction to
Microfluidics}}}}\ (\bibinfo  {publisher} {Oxford University Press},\
\bibinfo {year} {2005})\BibitemShut {NoStop}%
\bibitem [{\citenamefont {Jacobs}\ \emph {et~al.}(2008)\citenamefont {Jacobs},
\citenamefont {Seemann},\ and\ \citenamefont {Herminghaus}}]{Jacobs2008}%
\BibitemOpen
\bibfield  {author} {\bibinfo {author} {\bibfnamefont {K.}~\bibnamefont
{Jacobs}}, \bibinfo {author} {\bibfnamefont {R.}~\bibnamefont {Seemann}}, \
and\ \bibinfo {author} {\bibfnamefont {S.}~\bibnamefont {Herminghaus}},\ }in\
\href@noop {} {\emph {\bibinfo {booktitle} {Polym. Thin Film.}}},\ \bibinfo
{editor} {edited by\ \bibinfo {editor} {\bibfnamefont {O.~K.~C.}\
\bibnamefont {Tsui}}\ and\ \bibinfo {editor} {\bibfnamefont {T.~P.}\
\bibnamefont {Russell}}}\ (\bibinfo  {publisher} {World Scientific},\
\bibinfo {year} {2008})\ Chap.~\bibinfo {chapter} {10}, pp.\ \bibinfo {pages}
{243--265}\BibitemShut {NoStop}%
\bibitem [{\citenamefont {B\"{a}umchen}\ and\ \citenamefont
{Jacobs}(2010)}]{Baumchen2010b}%
\BibitemOpen
\bibfield  {author} {\bibinfo {author} {\bibfnamefont {O.}~\bibnamefont
{B\"{a}umchen}}\ and\ \bibinfo {author} {\bibfnamefont {K.}~\bibnamefont
{Jacobs}},\ }\href {\doibase 10.1088/0953-8984/22/3/033102} {\bibfield
{journal} {\bibinfo  {journal} {J. Phys. Condens. Matter}\ }\textbf {\bibinfo
{volume} {22}},\ \bibinfo {pages} {033102} (\bibinfo {year}
{2010})}\BibitemShut {NoStop}%
\bibitem [{\citenamefont {Hern\'andez-S\'anchez}\ \emph
{et~al.}(2012)\citenamefont {Hern\'andez-S\'anchez}, \citenamefont {Lubbers},
\citenamefont {Eddi},\ and\ \citenamefont {Snoeijer}}]{Hernandez2012}%
\BibitemOpen
\bibfield  {author} {\bibinfo {author} {\bibfnamefont {J.~F.}\ \bibnamefont
{Hern\'andez-S\'anchez}}, \bibinfo {author} {\bibfnamefont {L.~A.}\
\bibnamefont {Lubbers}}, \bibinfo {author} {\bibfnamefont {A.}~\bibnamefont
{Eddi}}, \ and\ \bibinfo {author} {\bibfnamefont {J.~H.}\ \bibnamefont
{Snoeijer}},\ }\href {\doibase 10.1103/PhysRevLett.109.184502} {\bibfield
{journal} {\bibinfo  {journal} {Phys. Rev. Lett.}\ }\textbf {\bibinfo
{volume} {109}},\ \bibinfo {pages} {184502} (\bibinfo {year}
{2012})}\BibitemShut {NoStop}%
\bibitem [{\citenamefont {Blossey}(2012)}]{Blossey2012}%
\BibitemOpen
\bibfield  {author} {\bibinfo {author} {\bibfnamefont {R.}~\bibnamefont
{Blossey}},\ }\href@noop {} {\emph {\bibinfo {title} {{Thin Liquid Films:
Dewetting and Polymer Flow}}}}\ (\bibinfo  {publisher} {Springer},\ \bibinfo
{year} {2012})\BibitemShut {NoStop}%
\bibitem [{\citenamefont {Oron}\ \emph {et~al.}(1997)\citenamefont {Oron},
\citenamefont {Davis},\ and\ \citenamefont {Bankoff}}]{Oron1997}%
\BibitemOpen
\bibfield  {author} {\bibinfo {author} {\bibfnamefont {A.}~\bibnamefont
{Oron}}, \bibinfo {author} {\bibfnamefont {S.}~\bibnamefont {Davis}}, \ and\
\bibinfo {author} {\bibfnamefont {S.}~\bibnamefont {Bankoff}},\ }\href
{http://journals.aps.org/rmp/abstract/10.1103/RevModPhys.69.931} {\bibfield
{journal} {\bibinfo  {journal} {Rev. Mod. Phys.}\ }\textbf {\bibinfo {volume}
{69}},\ \bibinfo {pages} {931} (\bibinfo {year} {1997})}\BibitemShut
{NoStop}%
\bibitem [{\citenamefont {Craster}\ and\ \citenamefont
{Matar}(2009)}]{Craster2009}%
\BibitemOpen
\bibfield  {author} {\bibinfo {author} {\bibfnamefont {R.}~\bibnamefont
{Craster}}\ and\ \bibinfo {author} {\bibfnamefont {O.}~\bibnamefont
{Matar}},\ }\href {\doibase 10.1103/RevModPhys.81.1131} {\bibfield  {journal}
{\bibinfo  {journal} {Rev. Mod. Phys.}\ }\textbf {\bibinfo {volume} {81}},\
\bibinfo {pages} {1131} (\bibinfo {year} {2009})}\BibitemShut {NoStop}%
\bibitem [{\citenamefont {Israelachvili}\ \emph {et~al.}(1988)\citenamefont
{Israelachvili}, \citenamefont {McGuiggan},\ and\ \citenamefont
{Homola}}]{Israelachvili1988}%
\BibitemOpen
\bibfield  {author} {\bibinfo {author} {\bibfnamefont {J.~N.}\ \bibnamefont
{Israelachvili}}, \bibinfo {author} {\bibfnamefont {P.~M.}\ \bibnamefont
{McGuiggan}}, \ and\ \bibinfo {author} {\bibfnamefont {A.~M.}\ \bibnamefont
{Homola}},\ }\href {\doibase 10.1126/science.240.4849.189} {\bibfield
{journal} {\bibinfo  {journal} {Science}\ }\textbf {\bibinfo {volume}
{240}},\ \bibinfo {pages} {189} (\bibinfo {year} {1988})}\BibitemShut
{NoStop}%
\bibitem [{\citenamefont {Villey}\ \emph {et~al.}(2013)\citenamefont {Villey},
\citenamefont {Martinot}, \citenamefont {Cottin-Bizonne}, \citenamefont
{Phaner-Goutorbe}, \citenamefont {L\'eger}, \citenamefont {Restagno},\ and\
\citenamefont {Charlaix}}]{Villey2013}%
\BibitemOpen
\bibfield  {author} {\bibinfo {author} {\bibfnamefont {R.}~\bibnamefont
{Villey}}, \bibinfo {author} {\bibfnamefont {E.}~\bibnamefont {Martinot}},
\bibinfo {author} {\bibfnamefont {C.}~\bibnamefont {Cottin-Bizonne}},
\bibinfo {author} {\bibfnamefont {M.}~\bibnamefont {Phaner-Goutorbe}},
\bibinfo {author} {\bibfnamefont {L.}~\bibnamefont {L\'eger}}, \bibinfo
{author} {\bibfnamefont {F.}~\bibnamefont {Restagno}}, \ and\ \bibinfo
{author} {\bibfnamefont {E.}~\bibnamefont {Charlaix}},\ }\href {\doibase
10.1103/PhysRevLett.111.215701} {\bibfield  {journal} {\bibinfo  {journal}
{Phys. Rev. Lett.}\ }\textbf {\bibinfo {volume} {111}},\ \bibinfo {pages}
{215701} (\bibinfo {year} {2013})}\BibitemShut {NoStop}%
\bibitem [{\citenamefont {Teichroeb}\ and\ \citenamefont
{Forrest}(2003)}]{Teichroeb2003}%
\BibitemOpen
\bibfield  {author} {\bibinfo {author} {\bibfnamefont {J.~H.}\ \bibnamefont
{Teichroeb}}\ and\ \bibinfo {author} {\bibfnamefont {J.~A.}\ \bibnamefont
{Forrest}},\ }\href {\doibase 10.1103/PhysRevLett.91.016104} {\bibfield
{journal} {\bibinfo  {journal} {Phys. Rev. Lett.}\ }\textbf {\bibinfo
{volume} {91}},\ \bibinfo {pages} {016104} (\bibinfo {year}
{2003})}\BibitemShut {NoStop}%
\bibitem [{\citenamefont {Schaffer}\ \emph {et~al.}(2000)\citenamefont
{Schaffer}, \citenamefont {Thurn-Albrecht}, \citenamefont {Russell},\ and\
\citenamefont {Steiner}}]{Schaffer2000}%
\BibitemOpen
\bibfield  {author} {\bibinfo {author} {\bibfnamefont {E.}~\bibnamefont
{Schaffer}}, \bibinfo {author} {\bibfnamefont {T.}~\bibnamefont
{Thurn-Albrecht}}, \bibinfo {author} {\bibfnamefont {T.}~\bibnamefont
{Russell}}, \ and\ \bibinfo {author} {\bibfnamefont {U.}~\bibnamefont
{Steiner}},\ }\href {\doibase 10.1038/35002540} {\bibfield  {journal}
{\bibinfo  {journal} {Nature}\ }\textbf {\bibinfo {volume} {403}},\ \bibinfo
{pages} {874} (\bibinfo {year} {2000})}\BibitemShut {NoStop}%
\bibitem [{\citenamefont {Morariu}\ \emph {et~al.}(2003)\citenamefont
{Morariu}, \citenamefont {Voicu}, \citenamefont {Sch\"{a}ffer}, \citenamefont
{Lin}, \citenamefont {Russell},\ and\ \citenamefont {Steiner}}]{Morariu2003}%
\BibitemOpen
\bibfield  {author} {\bibinfo {author} {\bibfnamefont {M.~D.}\ \bibnamefont
{Morariu}}, \bibinfo {author} {\bibfnamefont {N.~E.}\ \bibnamefont {Voicu}},
\bibinfo {author} {\bibfnamefont {E.}~\bibnamefont {Sch\"{a}ffer}}, \bibinfo
{author} {\bibfnamefont {Z.}~\bibnamefont {Lin}}, \bibinfo {author}
{\bibfnamefont {T.~P.}\ \bibnamefont {Russell}}, \ and\ \bibinfo {author}
{\bibfnamefont {U.}~\bibnamefont {Steiner}},\ }\href {\doibase
10.1038/nmat789} {\bibfield  {journal} {\bibinfo  {journal} {Nat. Mater.}\
}\textbf {\bibinfo {volume} {2}},\ \bibinfo {pages} {48} (\bibinfo {year}
{2003})}\BibitemShut {NoStop}%
\bibitem [{\citenamefont {Voicu}\ \emph {et~al.}(2006)\citenamefont {Voicu},
\citenamefont {Harkema},\ and\ \citenamefont {Steiner}}]{Voicu2006}%
\BibitemOpen
\bibfield  {author} {\bibinfo {author} {\bibfnamefont {N.}~\bibnamefont
{Voicu}}, \bibinfo {author} {\bibfnamefont {S.}~\bibnamefont {Harkema}}, \
and\ \bibinfo {author} {\bibfnamefont {U.}~\bibnamefont {Steiner}},\ }\href
{\doibase 10.1002/adfm.200500470} {\bibfield  {journal} {\bibinfo  {journal}
{Adv. Funct. Mater.}\ }\textbf {\bibinfo {volume} {16}},\ \bibinfo {pages}
{926} (\bibinfo {year} {2006})}\BibitemShut {NoStop}%
\bibitem [{\citenamefont {Srolovitz}\ and\ \citenamefont
{Safran}(1986)}]{Srolovitz1986}%
\BibitemOpen
\bibfield  {author} {\bibinfo {author} {\bibfnamefont {D.~J.}\ \bibnamefont
{Srolovitz}}\ and\ \bibinfo {author} {\bibfnamefont {S.~A.}\ \bibnamefont
{Safran}},\ }\href {\doibase 10.1063/1.337691} {\bibfield  {journal}
{\bibinfo  {journal} {J. Appl. Phys.}\ }\textbf {\bibinfo {volume} {60}},\
\bibinfo {pages} {255} (\bibinfo {year} {1986})}\BibitemShut {NoStop}%
\bibitem [{\citenamefont {Wyart}\ and\ \citenamefont
{Daillant}(1990)}]{Wyart1990}%
\BibitemOpen
\bibfield  {author} {\bibinfo {author} {\bibfnamefont {F.}~\bibnamefont
{Wyart}}\ and\ \bibinfo {author} {\bibfnamefont {J.}~\bibnamefont
{Daillant}},\ }\href
{http://www.nrcresearchpress.com/doi/abs/10.1139/p90-151} {\bibfield
{journal} {\bibinfo  {journal} {Can. J. Phys.}\ }\textbf {\bibinfo {volume}
{68}},\ \bibinfo {pages} {1084} (\bibinfo {year} {1990})}\BibitemShut
{NoStop}%
\bibitem [{\citenamefont {Seemann}\ \emph {et~al.}(2001)\citenamefont
{Seemann}, \citenamefont {Herminghaus},\ and\ \citenamefont
{Jacobs}}]{Seemann2001a}%
\BibitemOpen
\bibfield  {author} {\bibinfo {author} {\bibfnamefont {R.}~\bibnamefont
{Seemann}}, \bibinfo {author} {\bibfnamefont {S.}~\bibnamefont
{Herminghaus}}, \ and\ \bibinfo {author} {\bibfnamefont {K.}~\bibnamefont
{Jacobs}},\ }\href {\doibase 10.1103/PhysRevLett.87.196101} {\bibfield
{journal} {\bibinfo  {journal} {Phys. Rev. Lett.}\ }\textbf {\bibinfo
{volume} {87}},\ \bibinfo {pages} {196101} (\bibinfo {year}
{2001})}\BibitemShut {NoStop}%
\bibitem [{\citenamefont {Reiter}\ \emph {et~al.}(2005)\citenamefont {Reiter},
\citenamefont {Hamieh}, \citenamefont {Damman}, \citenamefont {Sclavons},
\citenamefont {Gabriele}, \citenamefont {Vilmin},\ and\ \citenamefont
{Rapha\"{e}l}}]{Reiter2005a}%
\BibitemOpen
\bibfield  {author} {\bibinfo {author} {\bibfnamefont {G.}~\bibnamefont
{Reiter}}, \bibinfo {author} {\bibfnamefont {M.}~\bibnamefont {Hamieh}},
\bibinfo {author} {\bibfnamefont {P.}~\bibnamefont {Damman}}, \bibinfo
{author} {\bibfnamefont {S.}~\bibnamefont {Sclavons}}, \bibinfo {author}
{\bibfnamefont {S.}~\bibnamefont {Gabriele}}, \bibinfo {author}
{\bibfnamefont {T.}~\bibnamefont {Vilmin}}, \ and\ \bibinfo {author}
{\bibfnamefont {E.}~\bibnamefont {Rapha\"{e}l}},\ }\href {\doibase
10.1038/nmat1484} {\bibfield  {journal} {\bibinfo  {journal} {Nat. Mater.}\
}\textbf {\bibinfo {volume} {4}},\ \bibinfo {pages} {754} (\bibinfo {year}
{2005})}\BibitemShut {NoStop}%
\bibitem [{\citenamefont {Chen}\ \emph {et~al.}(2012)\citenamefont {Chen},
\citenamefont {Li}, \citenamefont {Fang}, \citenamefont {Wu}, \citenamefont
{Wang}, \citenamefont {Ma}, \citenamefont {Ma}, \citenamefont {Shu},\ and\
\citenamefont {Peng}}]{Chen2012}%
\BibitemOpen
\bibfield  {author} {\bibinfo {author} {\bibfnamefont {X.-C.}\ \bibnamefont
{Chen}}, \bibinfo {author} {\bibfnamefont {H.-M.}\ \bibnamefont {Li}},
\bibinfo {author} {\bibfnamefont {F.}~\bibnamefont {Fang}}, \bibinfo {author}
{\bibfnamefont {Y.-W.}\ \bibnamefont {Wu}}, \bibinfo {author} {\bibfnamefont
{M.}~\bibnamefont {Wang}}, \bibinfo {author} {\bibfnamefont {G.-B.}\
\bibnamefont {Ma}}, \bibinfo {author} {\bibfnamefont {Y.-Q.}\ \bibnamefont
{Ma}}, \bibinfo {author} {\bibfnamefont {D.-J.}\ \bibnamefont {Shu}}, \ and\
\bibinfo {author} {\bibfnamefont {R.-W.}\ \bibnamefont {Peng}},\ }\href
{\doibase 10.1002/adma.201200089} {\bibfield  {journal} {\bibinfo  {journal}
{Adv. Mater.}\ }\textbf {\bibinfo {volume} {24}},\ \bibinfo {pages} {2637}
(\bibinfo {year} {2012})}\BibitemShut {NoStop}%
\bibitem [{\citenamefont {B\"{a}umchen}\ \emph {et~al.}(2009)\citenamefont
{B\"{a}umchen}, \citenamefont {Fetzer},\ and\ \citenamefont
{Jacobs}}]{Baumchen2009}%
\BibitemOpen
\bibfield  {author} {\bibinfo {author} {\bibfnamefont {O.}~\bibnamefont
{B\"{a}umchen}}, \bibinfo {author} {\bibfnamefont {R.}~\bibnamefont
{Fetzer}}, \ and\ \bibinfo {author} {\bibfnamefont {K.}~\bibnamefont
{Jacobs}},\ }\href {\doibase 10.1103/PhysRevLett.103.247801} {\bibfield
{journal} {\bibinfo  {journal} {Phys. Rev. Lett.}\ }\textbf {\bibinfo
{volume} {103}},\ \bibinfo {pages} {1} (\bibinfo {year} {2009})}\BibitemShut
{NoStop}%
\bibitem [{\citenamefont {Katzenstein}\ \emph {et~al.}(2012)\citenamefont
{Katzenstein}, \citenamefont {Janes}, \citenamefont {Cushen}, \citenamefont
{Hira}, \citenamefont {McGuffin}, \citenamefont {Prisco},\ and\ \citenamefont
{Ellison}}]{Katzenstein2012}%
\BibitemOpen
\bibfield  {author} {\bibinfo {author} {\bibfnamefont {J.~M.}\ \bibnamefont
{Katzenstein}}, \bibinfo {author} {\bibfnamefont {D.~W.}\ \bibnamefont
{Janes}}, \bibinfo {author} {\bibfnamefont {J.~D.}\ \bibnamefont {Cushen}},
\bibinfo {author} {\bibfnamefont {N.~B.}\ \bibnamefont {Hira}}, \bibinfo
{author} {\bibfnamefont {D.~L.}\ \bibnamefont {McGuffin}}, \bibinfo {author}
{\bibfnamefont {N.~A.}\ \bibnamefont {Prisco}}, \ and\ \bibinfo {author}
{\bibfnamefont {C.~J.}\ \bibnamefont {Ellison}},\ }\href {\doibase
10.1021/mz300400p} {\bibfield  {journal} {\bibinfo  {journal} {ACS Macro
Lett.}\ ,\ \bibinfo {pages} {1150}} (\bibinfo {year} {2012})}\BibitemShut
{NoStop}%
\bibitem [{\citenamefont {Kim}\ \emph {et~al.}(2014)\citenamefont {Kim},
\citenamefont {Janes}, \citenamefont {McGuffin},\ and\ \citenamefont
{Ellison}}]{BinKim2014}%
\BibitemOpen
\bibfield  {author} {\bibinfo {author} {\bibfnamefont {C.~B.}\ \bibnamefont
{Kim}}, \bibinfo {author} {\bibfnamefont {D.~W.}\ \bibnamefont {Janes}},
\bibinfo {author} {\bibfnamefont {D.~L.}\ \bibnamefont {McGuffin}}, \ and\
\bibinfo {author} {\bibfnamefont {C.~J.}\ \bibnamefont {Ellison}},\ }\href
{\doibase 10.1002/polb.23546} {\bibfield  {journal} {\bibinfo  {journal} {J.
Polym. Sci. Part B Polym. Phys.}\ }\textbf {\bibinfo {volume} {52}},\
\bibinfo {pages} {1195} (\bibinfo {year} {2014})}\BibitemShut {NoStop}%
\bibitem [{\citenamefont {Tsui}\ \emph {et~al.}(2008)\citenamefont {Tsui},
\citenamefont {Wang}, \citenamefont {Lee}, \citenamefont {Lam},\ and\
\citenamefont {Yang}}]{Tsui2008a}%
\BibitemOpen
\bibfield  {author} {\bibinfo {author} {\bibfnamefont {O.~K.~C.}\
\bibnamefont {Tsui}}, \bibinfo {author} {\bibfnamefont {Y.~J.}\ \bibnamefont
{Wang}}, \bibinfo {author} {\bibfnamefont {F.~K.}\ \bibnamefont {Lee}},
\bibinfo {author} {\bibfnamefont {C.-H.}\ \bibnamefont {Lam}}, \ and\
\bibinfo {author} {\bibfnamefont {Z.}~\bibnamefont {Yang}},\ }\href {\doibase
10.1021/ma702374g} {\bibfield  {journal} {\bibinfo  {journal}
{Macromolecules}\ }\textbf {\bibinfo {volume} {41}},\ \bibinfo {pages} {1465}
(\bibinfo {year} {2008})}\BibitemShut {NoStop}%
\bibitem [{\citenamefont {Fakhraai}\ and\ \citenamefont
{Forrest}(2008)}]{Fakhraai2008}%
\BibitemOpen
\bibfield  {author} {\bibinfo {author} {\bibfnamefont {Z.}~\bibnamefont
{Fakhraai}}\ and\ \bibinfo {author} {\bibfnamefont {J.~A.}\ \bibnamefont
{Forrest}},\ }\href {\doibase 10.1126/science.1151205} {\bibfield  {journal}
{\bibinfo  {journal} {Science}\ }\textbf {\bibinfo {volume} {319}},\ \bibinfo
{pages} {600} (\bibinfo {year} {2008})}\BibitemShut {NoStop}%
\bibitem [{\citenamefont {Leveder}\ \emph {et~al.}(2008)\citenamefont
{Leveder}, \citenamefont {Landis},\ and\ \citenamefont
{Davoust}}]{Leveder2008}%
\BibitemOpen
\bibfield  {author} {\bibinfo {author} {\bibfnamefont {T.}~\bibnamefont
{Leveder}}, \bibinfo {author} {\bibfnamefont {S.}~\bibnamefont {Landis}}, \
and\ \bibinfo {author} {\bibfnamefont {L.}~\bibnamefont {Davoust}},\ }\href
{\doibase 10.1063/1.2828986} {\bibfield  {journal} {\bibinfo  {journal}
{Appl. Phys. Lett.}\ }\textbf {\bibinfo {volume} {92}},\ \bibinfo {pages}
{013107} (\bibinfo {year} {2008})}\BibitemShut {NoStop}%
\bibitem [{\citenamefont {Yang}\ \emph {et~al.}(2010)\citenamefont {Yang},
\citenamefont {Fujii}, \citenamefont {Lee}, \citenamefont {Lam},\ and\
\citenamefont {Tsui}}]{Yang2010}%
\BibitemOpen
\bibfield  {author} {\bibinfo {author} {\bibfnamefont {Z.}~\bibnamefont
{Yang}}, \bibinfo {author} {\bibfnamefont {Y.}~\bibnamefont {Fujii}},
\bibinfo {author} {\bibfnamefont {F.~K.}\ \bibnamefont {Lee}}, \bibinfo
{author} {\bibfnamefont {C.-H.}\ \bibnamefont {Lam}}, \ and\ \bibinfo
{author} {\bibfnamefont {O.~K.~C.}\ \bibnamefont {Tsui}},\ }\href {\doibase
10.1126/science.1184394} {\bibfield  {journal} {\bibinfo  {journal}
{Science}\ }\textbf {\bibinfo {volume} {328}},\ \bibinfo {pages} {1676}
(\bibinfo {year} {2010})}\BibitemShut {NoStop}%
\bibitem [{\citenamefont {McGraw}\ \emph {et~al.}(2012)\citenamefont {McGraw},
\citenamefont {Salez}, \citenamefont {B\"{a}umchen}, \citenamefont
{Rapha\"{e}l},\ and\ \citenamefont {Dalnoki-Veress}}]{McGraw2012}%
\BibitemOpen
\bibfield  {author} {\bibinfo {author} {\bibfnamefont {J.~D.}\ \bibnamefont
{McGraw}}, \bibinfo {author} {\bibfnamefont {T.}~\bibnamefont {Salez}},
\bibinfo {author} {\bibfnamefont {O.}~\bibnamefont {B\"{a}umchen}}, \bibinfo
{author} {\bibfnamefont {E.}~\bibnamefont {Rapha\"{e}l}}, \ and\ \bibinfo
{author} {\bibfnamefont {K.}~\bibnamefont {Dalnoki-Veress}},\ }\href
{\doibase 10.1103/PhysRevLett.109.128303} {\bibfield  {journal} {\bibinfo
{journal} {Phys. Rev. Lett.}\ }\textbf {\bibinfo {volume} {109}},\ \bibinfo
{pages} {128303} (\bibinfo {year} {2012})}\BibitemShut {NoStop}%
\bibitem [{\citenamefont {Rognin}\ \emph {et~al.}(2012)\citenamefont {Rognin},
\citenamefont {Landis},\ and\ \citenamefont {Davoust}}]{Rognin2012}%
\BibitemOpen
\bibfield  {author} {\bibinfo {author} {\bibfnamefont {E.}~\bibnamefont
{Rognin}}, \bibinfo {author} {\bibfnamefont {S.}~\bibnamefont {Landis}}, \
and\ \bibinfo {author} {\bibfnamefont {L.}~\bibnamefont {Davoust}},\ }\href
{\doibase 10.1116/1.3664088} {\bibfield  {journal} {\bibinfo  {journal} {J.
Vac. Sci. Technol. B Microelectron. Nanom. Struct.}\ }\textbf {\bibinfo
{volume} {30}},\ \bibinfo {pages} {011602} (\bibinfo {year}
{2012})}\BibitemShut {NoStop}%
\bibitem [{\citenamefont {Chai}\ \emph {et~al.}(2014)\citenamefont {Chai},
\citenamefont {Salez}, \citenamefont {McGraw}, \citenamefont {Benzaquen},
\citenamefont {Dalnoki-Veress}, \citenamefont {Rapha\"{e}l},\ and\
\citenamefont {Forrest}}]{Chai2014}%
\BibitemOpen
\bibfield  {author} {\bibinfo {author} {\bibfnamefont {Y.}~\bibnamefont
{Chai}}, \bibinfo {author} {\bibfnamefont {T.}~\bibnamefont {Salez}},
\bibinfo {author} {\bibfnamefont {J.~D.}\ \bibnamefont {McGraw}}, \bibinfo
{author} {\bibfnamefont {M.}~\bibnamefont {Benzaquen}}, \bibinfo {author}
{\bibfnamefont {K.}~\bibnamefont {Dalnoki-Veress}}, \bibinfo {author}
{\bibfnamefont {E.}~\bibnamefont {Rapha\"{e}l}}, \ and\ \bibinfo {author}
{\bibfnamefont {J.~A.}\ \bibnamefont {Forrest}},\ }\href {\doibase
10.1126/science.1244845} {\bibfield  {journal} {\bibinfo  {journal}
{Science}\ }\textbf {\bibinfo {volume} {343}},\ \bibinfo {pages} {994}
(\bibinfo {year} {2014})}\BibitemShut {NoStop}%
\bibitem [{\citenamefont {Rutgers}\ \emph {et~al.}(1996)\citenamefont
{Rutgers}, \citenamefont {Wu}, \citenamefont {Bhagavatula}, \citenamefont
{Petersen},\ and\ \citenamefont {Goldburg}}]{Rutgers1996}%
\BibitemOpen
\bibfield  {author} {\bibinfo {author} {\bibfnamefont {M.~A.}\ \bibnamefont
{Rutgers}}, \bibinfo {author} {\bibfnamefont {X.-l.}\ \bibnamefont {Wu}},
\bibinfo {author} {\bibfnamefont {R.}~\bibnamefont {Bhagavatula}}, \bibinfo
{author} {\bibfnamefont {A.~A.}\ \bibnamefont {Petersen}}, \ and\ \bibinfo
{author} {\bibfnamefont {W.~I.}\ \bibnamefont {Goldburg}},\ }\href {\doibase
10.1063/1.869105} {\bibfield  {journal} {\bibinfo  {journal} {Phys. Fluids}\
}\textbf {\bibinfo {volume} {8}},\ \bibinfo {pages} {2847} (\bibinfo {year}
{1996})}\BibitemShut {NoStop}%
\bibitem [{\citenamefont {Horv\'{a}th}\ \emph {et~al.}(2000)\citenamefont
{Horv\'{a}th}, \citenamefont {Cressman}, \citenamefont {Goldburg},\ and\
\citenamefont {Wu}}]{Horvath2000}%
\BibitemOpen
\bibfield  {author} {\bibinfo {author} {\bibfnamefont {V.}~\bibnamefont
{Horv\'{a}th}}, \bibinfo {author} {\bibfnamefont {J.}~\bibnamefont
{Cressman}}, \bibinfo {author} {\bibfnamefont {W.}~\bibnamefont {Goldburg}},
\ and\ \bibinfo {author} {\bibfnamefont {X.}~\bibnamefont {Wu}},\ }\href
{\doibase 10.1103/PhysRevE.61.R4702} {\bibfield  {journal} {\bibinfo
{journal} {Phys. Rev. E}\ }\textbf {\bibinfo {volume} {61}},\ \bibinfo
{pages} {R4702} (\bibinfo {year} {2000})}\BibitemShut {NoStop}%
\bibitem [{\citenamefont {Aradian}\ \emph {et~al.}(2001)\citenamefont
{Aradian}, \citenamefont {Rapha\"{e}l},\ and\ \citenamefont
{de Gennes}}]{Aradian2001}%
\BibitemOpen
\bibfield  {author} {\bibinfo {author} {\bibfnamefont {A.}~\bibnamefont
{Aradian}}, \bibinfo {author} {\bibfnamefont {E.}~\bibnamefont {Rapha\"{e}l}}, \
and\ \bibinfo {author} {\bibfnamefont {P.-G.}~\bibnamefont {de Gennes}},\
}\href@noop {} {\bibfield  {journal} {\bibinfo  {journal} {Europhysics Letters}\
}\textbf {\bibinfo {volume} {{55}}},\ \bibinfo {pages}{834} (\bibinfo {year}
{2001})}\BibitemShut {NoStop}%
\bibitem [{\citenamefont {Georgiev}\ and\ \citenamefont
{Vorobieff}(2002)}]{Georgiev2002}%
\BibitemOpen
\bibfield  {author} {\bibinfo {author} {\bibfnamefont {D.}~\bibnamefont
{Georgiev}}\ and\ \bibinfo {author} {\bibfnamefont {P.}~\bibnamefont
{Vorobieff}},\ }\href {\doibase 10.1063/1.1446040} {\bibfield  {journal}
{\bibinfo  {journal} {Rev. Sci. Instrum.}\ }\textbf {\bibinfo {volume}
{73}},\ \bibinfo {pages} {1177} (\bibinfo {year} {2002})}\BibitemShut
{NoStop}%
\bibitem [{\citenamefont {Diamant}(2009)}]{Diamant2009}%
\BibitemOpen
\bibfield  {author} {\bibinfo {author} {\bibfnamefont {H.}~\bibnamefont
{Diamant}},\ }\href {\doibase 10.1143/JPSJ.78.041002} {\bibfield  {journal}
{\bibinfo  {journal} {J. Phys. Soc. Japan}\ }\textbf {\bibinfo {volume}
{78}},\ \bibinfo {pages} {041002} (\bibinfo {year} {2009})}\BibitemShut
{NoStop}%
\bibitem [{\citenamefont {Seiwert}\ \emph {et~al.}(2013)\citenamefont
{Seiwert}, \citenamefont {Monloubou}, \citenamefont {Dollet},\ and\
\citenamefont {Cantat}}]{Seiwert2013}%
\BibitemOpen
\bibfield  {author} {\bibinfo {author} {\bibfnamefont {J.}~\bibnamefont
{Seiwert}}, \bibinfo {author} {\bibfnamefont {M.}~\bibnamefont {Monloubou}},
\bibinfo {author} {\bibfnamefont {B.}~\bibnamefont {Dollet}}, \ and\ \bibinfo
{author} {\bibfnamefont {I.}~\bibnamefont {Cantat}},\ }\href {\doibase
10.1103/PhysRevLett.111.094501} {\bibfield  {journal} {\bibinfo  {journal}
{Phys. Rev. Lett.}\ }\textbf {\bibinfo {volume} {111}},\ \bibinfo {pages}
{094501} (\bibinfo {year} {2013})}\BibitemShut {NoStop}%
\bibitem [{\citenamefont {Huang}\ \emph {et~al.}(2004)\citenamefont {Huang},
\citenamefont {Wen}, \citenamefont {Lee},\ and\ \citenamefont
{Tsai}}]{Huang2013}%
\BibitemOpen
\bibfield  {author} {\bibinfo {author} {\bibfnamefont {M.~J.}\ \bibnamefont
{Huang}}, \bibinfo {author} {\bibfnamefont {C.~Y.}\ \bibnamefont {Wen}},
\bibinfo {author} {\bibfnamefont {I.~C.}\ \bibnamefont {Lee}}, \ and\
\bibinfo {author} {\bibfnamefont {C.~H.}\ \bibnamefont {Tsai}},\ }\href
{\doibase 10.1063/1.1802131} {\bibfield  {journal} {\bibinfo  {journal}
{Phys. Fluids}\ }\textbf {\bibinfo {volume} {16}},\ \bibinfo {pages} {3975}
(\bibinfo {year} {2004})}\BibitemShut {NoStop}%
\bibitem [{\citenamefont {Harth}\ \emph {et~al.}(2011)\citenamefont {Harth},
\citenamefont {Eremin},\ and\ \citenamefont {Stannarius}}]{Harth2011}%
\BibitemOpen
\bibfield  {author} {\bibinfo {author} {\bibfnamefont {K.}~\bibnamefont
{Harth}}, \bibinfo {author} {\bibfnamefont {A.}~\bibnamefont {Eremin}}, \
and\ \bibinfo {author} {\bibfnamefont {R.}~\bibnamefont {Stannarius}},\
}\href {\doibase 10.1039/c0sm01040e} {\bibfield  {journal} {\bibinfo
{journal} {Soft Matter}\ }\textbf {\bibinfo {volume} {7}},\ \bibinfo {pages}
{2858} (\bibinfo {year} {2011})}\BibitemShut {NoStop}%
\bibitem [{\citenamefont {Eremin}\ \emph {et~al.}(2006)\citenamefont {Eremin},
\citenamefont {Bohley},\ and\ \citenamefont {Stannarius}}]{Eremin2006}%
\BibitemOpen
\bibfield  {author} {\bibinfo {author} {\bibfnamefont {A.}~\bibnamefont
{Eremin}}, \bibinfo {author} {\bibfnamefont {C.}~\bibnamefont {Bohley}}, \
and\ \bibinfo {author} {\bibfnamefont {R.}~\bibnamefont {Stannarius}},\
}\href {\doibase 10.1103/PhysRevE.74.040701} {\bibfield  {journal} {\bibinfo
{journal} {Phys. Rev. E - Stat. Nonlinear, Soft Matter Phys.}\ }\textbf
{\bibinfo {volume} {74}},\ \bibinfo {pages} {1} (\bibinfo {year}
{2006})}\BibitemShut {NoStop}%
\bibitem [{\citenamefont {Stannarius}\ \emph {et~al.}(2006)\citenamefont
{Stannarius}, \citenamefont {Bohley},\ and\ \citenamefont
{Eremin}}]{Stannarius2006}%
\BibitemOpen
\bibfield  {author} {\bibinfo {author} {\bibfnamefont {R.}~\bibnamefont
{Stannarius}}, \bibinfo {author} {\bibfnamefont {C.}~\bibnamefont {Bohley}},
\ and\ \bibinfo {author} {\bibfnamefont {A.}~\bibnamefont {Eremin}},\ }\href
{\doibase 10.1103/PhysRevLett.97.097802} {\bibfield  {journal} {\bibinfo
{journal} {Phys. Rev. Lett.}\ }\textbf {\bibinfo {volume} {97}},\ \bibinfo
{pages} {1} (\bibinfo {year} {2006})}\BibitemShut {NoStop}%
\bibitem [{\citenamefont {Keddie}\ \emph {et~al.}(1994)\citenamefont {Keddie},
\citenamefont {Jones},\ and\ \citenamefont {Cory}}]{Keddie1994}%
\BibitemOpen
\bibfield  {author} {\bibinfo {author} {\bibfnamefont {J.~L.}\ \bibnamefont
{Keddie}}, \bibinfo {author} {\bibfnamefont {R.~A.~L.}\ \bibnamefont
{Jones}}, \ and\ \bibinfo {author} {\bibfnamefont {R.~A.}\ \bibnamefont
{Cory}},\ }\href@noop {} {\bibfield  {journal} {\bibinfo  {journal} {EPL
(Europhysics Letters)}\ }\textbf {\bibinfo {volume} {27}},\ \bibinfo {pages}
{59} (\bibinfo {year} {1994})}\BibitemShut {NoStop}%
\bibitem [{\citenamefont {Forrest}\ \emph {et~al.}(1996)\citenamefont
{Forrest}, \citenamefont {Dalnoki-Veress}, \citenamefont {Stevens},\ and\
\citenamefont {Dutcher}}]{Forrest1996}%
\BibitemOpen
\bibfield  {author} {\bibinfo {author} {\bibfnamefont {J.~A.}\ \bibnamefont
{Forrest}}, \bibinfo {author} {\bibfnamefont {K.}~\bibnamefont
{Dalnoki-Veress}}, \bibinfo {author} {\bibfnamefont {J.~R.}\ \bibnamefont
{Stevens}}, \ and\ \bibinfo {author} {\bibfnamefont {J.~R.}\ \bibnamefont
{Dutcher}},\ }\href@noop {} {\bibfield  {journal} {\bibinfo  {journal}
{Physical Review Letters}\ }\textbf {\bibinfo {volume} {77}},\ \bibinfo
{pages} {2002} (\bibinfo {year} {1996})}\BibitemShut {NoStop}%
\bibitem [{\citenamefont {Silberberg}(1982)}]{Silberberg1982}%
\BibitemOpen
\bibfield  {author} {\bibinfo {author} {\bibfnamefont {A.}~\bibnamefont
{Silberberg}},\ }\href@noop {} {\bibfield  {journal} {\bibinfo  {journal} {J.
Colloid Int. Sci.}\ }\textbf {\bibinfo {volume} {{\bf90}}},\ \bibinfo {pages}{86} (\bibinfo
{year} {1982})}\BibitemShut {NoStop}%
\bibitem [{\citenamefont {Brochard~Wyart}\ and\ \citenamefont
{de~Gennes}(2000)}]{Brochard2000}%
\BibitemOpen
\bibfield  {author} {\bibinfo {author} {\bibfnamefont {F.}~\bibnamefont
{Brochard~Wyart}}\ and\ \bibinfo {author} {\bibfnamefont {P.-G.}\
\bibnamefont {de~Gennes}},\ }\href@noop {} {\bibfield  {journal} {\bibinfo
{journal} {Eur. Phys. J. E}\ }\textbf {\bibinfo {volume} {{\bf1}}},\ \bibinfo {pages}{93}
(\bibinfo {year} {2000})}\BibitemShut {NoStop}%
\bibitem [{\citenamefont {Bodiguel}\ and\ \citenamefont
{Fretigny}(2006)}]{Bodiguel2006}%
\BibitemOpen
\bibfield  {author} {\bibinfo {author} {\bibfnamefont {H.}~\bibnamefont
{Bodiguel}}\ and\ \bibinfo {author} {\bibfnamefont {C.}~\bibnamefont
{Fretigny}},\ }\href@noop {} {\bibfield  {journal} {\bibinfo  {journal}
{Phys. Rev. Lett.}\ }\textbf {\bibinfo {volume} {{97}}},\ \bibinfo {pages}{266105} (\bibinfo
{year} {2006})}\BibitemShut {NoStop}%
\bibitem [{\citenamefont {Shin}\ \emph {et~al.}(2007)\citenamefont {Shin},
\citenamefont {Obukhov}, \citenamefont {Chen}, \citenamefont {Huh},
\citenamefont {Hwang}, \citenamefont {Mok}, \citenamefont {Dobriyal},
\citenamefont {Thiyagarajan},\ and\ \citenamefont {Russell}}]{Shin2007}%
\BibitemOpen
\bibfield  {author} {\bibinfo {author} {\bibfnamefont {K.}~\bibnamefont
{Shin}}, \bibinfo {author} {\bibfnamefont {S.}~\bibnamefont {Obukhov}},
\bibinfo {author} {\bibfnamefont {J.-T.}\ \bibnamefont {Chen}}, \bibinfo
{author} {\bibfnamefont {J.}~\bibnamefont {Huh}}, \bibinfo {author}
{\bibfnamefont {Y.}~\bibnamefont {Hwang}}, \bibinfo {author} {\bibfnamefont
{S.}~\bibnamefont {Mok}}, \bibinfo {author} {\bibfnamefont {P.}~\bibnamefont
{Dobriyal}}, \bibinfo {author} {\bibfnamefont {P.}~\bibnamefont
{Thiyagarajan}}, \ and\ \bibinfo {author} {\bibfnamefont {T.}~\bibnamefont
{Russell}},\ }\href@noop {} {\bibfield  {journal} {\bibinfo  {journal}
{Nature Materials}\ }\textbf {\bibinfo {volume} {{6}}},\ \bibinfo {pages}{961} (\bibinfo
{year} {2007})}\BibitemShut {NoStop}%
\bibitem [{\citenamefont {{Dalnoki-Veress, K.}}\ \emph
{et~al.}(2000)\citenamefont {{Dalnoki-Veress, K.}}, \citenamefont {{Forrest,
J. A.}}, \citenamefont {{de Gennes, P. G.}},\ and\ \citenamefont {{Dutcher,
J. R.}}}]{Dalnoki2000}%
\BibitemOpen
\bibfield  {author} {\bibinfo {author} {\bibnamefont {{Dalnoki-Veress, K.}}},
\bibinfo {author} {\bibnamefont {{Forrest, J. A.}}}, \bibinfo {author}
{\bibnamefont {{de Gennes, P. G.}}}, \ and\ \bibinfo {author} {\bibnamefont
{{Dutcher, J. R.}}},\ }\href@noop {} {\bibfield  {journal} {\bibinfo
{journal} {J. Phys. IV France}\ }\textbf {\bibinfo {volume} {10}},\ \bibinfo
{pages} {Pr7} (\bibinfo {year} {2000})}\BibitemShut {NoStop}%
\bibitem [{\citenamefont {Si}\ \emph {et~al.}(2005)\citenamefont {Si},
\citenamefont {Massa}, \citenamefont {Dalnoki-Veress}, \citenamefont
{Brown},\ and\ \citenamefont {Jones}}]{Si2005}%
\BibitemOpen
\bibfield  {author} {\bibinfo {author} {\bibfnamefont {L.}~\bibnamefont
{Si}}, \bibinfo {author} {\bibfnamefont {M.~V.}\ \bibnamefont {Massa}},
\bibinfo {author} {\bibfnamefont {K.}~\bibnamefont {Dalnoki-Veress}},
\bibinfo {author} {\bibfnamefont {H.~R.}\ \bibnamefont {Brown}}, \ and\
\bibinfo {author} {\bibfnamefont {R.~A.~L.}\ \bibnamefont {Jones}},\
}\href@noop {} {\bibfield  {journal} {\bibinfo  {journal} {Phys. Rev. Lett.}\
}\textbf {\bibinfo {volume} {{94}}},\ \bibinfo {pages}{127801} (\bibinfo {year}
{2005})}\BibitemShut {NoStop}%
\bibitem [{\citenamefont {Eggers}\ (1997)\citenamefont
{Eggers}}]{Eggers1997}%
\BibitemOpen
\bibfield  {author} {\bibinfo {author} {\bibfnamefont {J.}~\bibnamefont
{Eggers}},\ } {\bibfield  {journal} {\bibinfo
{journal} {Reviews of Modern Physics}\ }\textbf {\bibinfo {volume} {69}},\ \bibinfo
{pages} {865} (\bibinfo {year} {1997})}\BibitemShut {NoStop}%
\bibitem [{\citenamefont {Kargupta}\ \emph {et~al.}(2004)\citenamefont
{Kargupta}, \citenamefont {Sharma},\ and\ \citenamefont
{Khanna}}]{Kargupta2004}%
\BibitemOpen
\bibfield  {author} {\bibinfo {author} {\bibfnamefont {K.}~\bibnamefont
{Kargupta}}, \bibinfo {author} {\bibfnamefont {A.}~\bibnamefont {Sharma}}, \
and\ \bibinfo {author} {\bibfnamefont {R.}~\bibnamefont {Khanna}},\
}\href@noop {} {\bibfield  {journal} {\bibinfo  {journal} {Langmuir}\
}\textbf {\bibinfo {volume} {{20}}},\ \bibinfo {pages}{244} (\bibinfo {year}
{2004})}\BibitemShut {NoStop}%
\bibitem [{\citenamefont {M\"unch}\ \emph {et~al.}(2005)\citenamefont
{M\"unch}, \citenamefont {Wagner},\ and\ \citenamefont
{Witelski}}]{Munch2005}%
\BibitemOpen
\bibfield  {author} {\bibinfo {author} {\bibfnamefont {A.}~\bibnamefont
{M\"unch}}, \bibinfo {author} {\bibfnamefont {B.}~\bibnamefont {Wagner}}, \
and\ \bibinfo {author} {\bibfnamefont {T.~P.}\ \bibnamefont {Witelski}},\
}\href@noop {} {\bibfield  {journal} {\bibinfo  {journal} {Jour. Eng. Math.}\
}\textbf {\bibinfo {volume} {{\bf53}}},\ \bibinfo {pages} {359} (\bibinfo
{year} {2005})}\BibitemShut {NoStop}%
\bibitem [{\citenamefont {Howell}\ \emph {et~al.}(2005)\citenamefont
{Howell},\ and\ \citenamefont
{Stone}}]{Howell2005}%
\BibitemOpen
\bibfield  {author} {\bibinfo {author} {\bibfnamefont {P. D.}~\bibnamefont
{Howell}}, \
and\ \bibinfo {author} {\bibfnamefont {H. A.}~\bibnamefont {Stone}},\
}\href@noop {} {\bibfield  {journal} {\bibinfo  {journal} {European Journal of Applied Mathematics}}, \bibinfo {pages}{569} (\bibinfo {year}
{2005})}\BibitemShut {NoStop}%
\bibitem [{\citenamefont {Davidovitch}\ \emph {et~al.}(2005)\citenamefont
{Davidovitch},\  {Moro},\ and\ \citenamefont
{Stone}}]{Davidovitch2005}%
\BibitemOpen
\bibfield  {author} {\bibinfo {author} {\bibfnamefont {B.}~\bibnamefont
{Davidovitch}}, \ \bibinfo {author} {\bibfnamefont {E.}~\bibnamefont
{Moro}}, \
and\ \bibinfo {author} {\bibfnamefont {H. A.}~\bibnamefont {Stone}},\
}\href@noop {} {\bibfield  {journal} {\bibinfo  {journal} {Physical Review Letters}} \textbf {\bibinfo {volume} {95}}, \bibinfo {pages}{244505} (\bibinfo {year}
{2005})}\BibitemShut {NoStop}%
\bibitem [{\citenamefont {Gr{\"{u}}n}\ \emph {et~al.}(2006)\citenamefont
{Gr{\"{u}}n}, \citenamefont {Mecke},\ and\ \citenamefont
{Rauscher}}]{Grun2006}%
\BibitemOpen
\bibfield  {author} {\bibinfo {author} {\bibfnamefont {G.}~\bibnamefont
{Gr{\"{u}}n}}, \bibinfo {author} {\bibfnamefont {K.}~\bibnamefont {Mecke}}, \
and\ \bibinfo {author} {\bibfnamefont {M.}~\bibnamefont {Rauscher}},\
}\href@noop {} {\bibfield  {journal} {\bibinfo  {journal} {J. Stat. Phys.}\
}\textbf {\bibinfo {volume} {122}},\ \bibinfo {pages} {1261} (\bibinfo {year}
{2006})}\BibitemShut {NoStop}%
\bibitem [{\citenamefont {Hennequin}\ \emph {et~al.}(2006)\citenamefont
{Hennequin},\ {Aarts},\ {van der Wiel},\ {Wegdam},\ {Eggers},\ {Lekkerkerker},\and\ \citenamefont
{Bonn}}]{Hennequin2006}%
\BibitemOpen
\bibfield  {author} {\bibinfo {author} {\bibfnamefont {Y.}~\bibnamefont
{Hennequin}}, \ \bibinfo {author} {\bibfnamefont {D. G. A. L.}~\bibnamefont
{Aarts}}, \ \bibinfo {author} {\bibfnamefont {J. H.}~\bibnamefont
{van der Wiel}}, \ \bibinfo {author} {\bibfnamefont {G.}~\bibnamefont
{Wegdam}}, \ \bibinfo {author} {\bibfnamefont {J.}~\bibnamefont
{Eggers}},\ \bibinfo {author} {\bibfnamefont {H. N. W.}~\bibnamefont
{Lekkerkerker}},\
and\ \bibinfo {author} {\bibfnamefont {D.}~\bibnamefont {Bonn}},\
}\href@noop {} {\bibfield  {journal} {\bibinfo  {journal} {Physical Review Letters}} \textbf {\bibinfo {volume} {97}}, \bibinfo {pages}{244502} (\bibinfo {year}
{2006})}\BibitemShut {NoStop}%
\bibitem [{\citenamefont {McGraw}\ \emph {et~al.}(2011)\citenamefont {McGraw},
\citenamefont {Jago},\ and\ \citenamefont {Dalnoki-Veress}}]{McGraw2011}%
\BibitemOpen
\bibfield  {author} {\bibinfo {author} {\bibfnamefont {J.~D.}\ \bibnamefont
{McGraw}}, \bibinfo {author} {\bibfnamefont {N.~M.}\ \bibnamefont {Jago}}, \
and\ \bibinfo {author} {\bibfnamefont {K.}~\bibnamefont {Dalnoki-Veress}},\
}\href {\doibase 10.1039/c1sm05261f} {\bibfield  {journal} {\bibinfo
{journal} {Soft Matter}\ }\textbf {\bibinfo {volume} {7}},\ \bibinfo {pages}
{7832} (\bibinfo {year} {2011})}\BibitemShut {NoStop}%
\bibitem [{\citenamefont {B\"{a}umchen}\ \emph {et~al.}(2013)\citenamefont
{B\"{a}umchen}, \citenamefont {Benzaquen}, \citenamefont {Salez},
\citenamefont {McGraw}, \citenamefont {Backholm}, \citenamefont {Fowler},
\citenamefont {Rapha\"{e}l},\ and\ \citenamefont
{Dalnoki-Veress}}]{Baumchen2013}%
\BibitemOpen
\bibfield  {author} {\bibinfo {author} {\bibfnamefont {O.}~\bibnamefont
{B\"{a}umchen}}, \bibinfo {author} {\bibfnamefont {M.}~\bibnamefont
{Benzaquen}}, \bibinfo {author} {\bibfnamefont {T.}~\bibnamefont {Salez}},
\bibinfo {author} {\bibfnamefont {J.~D.}\ \bibnamefont {McGraw}}, \bibinfo
{author} {\bibfnamefont {M.}~\bibnamefont {Backholm}}, \bibinfo {author}
{\bibfnamefont {P.}~\bibnamefont {Fowler}}, \bibinfo {author} {\bibfnamefont
{E.}~\bibnamefont {Rapha\"{e}l}}, \ and\ \bibinfo {author} {\bibfnamefont
{K.}~\bibnamefont {Dalnoki-Veress}},\ }\href {\doibase
10.1103/PhysRevE.88.035001} {\bibfield  {journal} {\bibinfo  {journal} {Phys.
Rev. E}\ }\textbf {\bibinfo {volume} {88}},\ \bibinfo {pages} {035001}
(\bibinfo {year} {2013})}\BibitemShut {NoStop}%
\bibitem [{\citenamefont {Barenblatt}(1996)}]{Barenblatt1996}%
\BibitemOpen
\bibfield  {author} {\bibinfo {author} {\bibfnamefont {G.~I.}\ \bibnamefont
{Barenblatt}},\ }\href@noop {} {\emph {\bibinfo {title} {Scaling,
{S}elf-similarity, and {I}ntermediate {A}symptotics}}}\ (\bibinfo  {publisher}
{Cambridge University Press, Cambridge},\ \bibinfo {year} {1996})\BibitemShut
{NoStop}%
\bibitem [{\citenamefont {Salez}\ \emph {et~al.}(2012)\citenamefont {Salez},
\citenamefont {McGraw}, \citenamefont {Cormier}, \citenamefont
{B\"{a}umchen}, \citenamefont {Dalnoki-Veress},\ and\ \citenamefont
{Rapha\"{e}l}}]{Salez2012a}%
\BibitemOpen
\bibfield  {author} {\bibinfo {author} {\bibfnamefont {T.}~\bibnamefont
{Salez}}, \bibinfo {author} {\bibfnamefont {J.~D.}\ \bibnamefont {McGraw}},
\bibinfo {author} {\bibfnamefont {S.~L.}\ \bibnamefont {Cormier}}, \bibinfo
{author} {\bibfnamefont {O.}~\bibnamefont {B\"{a}umchen}}, \bibinfo {author}
{\bibfnamefont {K.}~\bibnamefont {Dalnoki-Veress}}, \ and\ \bibinfo {author}
{\bibfnamefont {E.}~\bibnamefont {Rapha\"{e}l}},\ }\href {\doibase
10.1140/epje/i2012-12114-x} {\bibfield  {journal} {\bibinfo  {journal} {Eur.
Phys. J. E}\ }\textbf {\bibinfo {volume} {35}},\ \bibinfo {pages} {114}
(\bibinfo {year} {2012})}\BibitemShut {NoStop}%
\end{thebibliography}
\end{document}